\documentclass[conference]{IEEEtran}

\usepackage[ruled,vlined]{algorithm2e}
\usepackage{graphicx}
\usepackage[usenames,dvipsnames]{xcolor}
\usepackage{numprint} 
\usepackage{subfig}
\usepackage{tabularx}
\usepackage{comment}
\input epsf
\usepackage{graphicx}
\begin{document}

\title{\LARGE A Deep Learning-based Penetration Testing Framework for Vulnerability Identification in Internet of Things  Environments}

\author{\authorblockN{Nickolaos Koroniotis \IEEEauthorrefmark{1}\IEEEauthorrefmark{2}, Nour Moustafa\IEEEauthorrefmark{1}\IEEEauthorrefmark{2}, Benjamin Turnbull\IEEEauthorrefmark{1}\IEEEauthorrefmark{2}, Francesco Schiliro\IEEEauthorrefmark{1}\IEEEauthorrefmark{3},\\ Praveen Gauravaram \IEEEauthorrefmark{1}\IEEEauthorrefmark{4}, Helge Janicke\IEEEauthorrefmark{1}}

\authorblockA{\authorrefmark{1}Cyber Security Cooperative Research Centre (CSCRC), Perth, WA 6027, Australia}
\authorblockA{\authorrefmark{2}School of Engineering and Information Technology, University of New South Wales at ADFA,\\ Canberra, ACT 2612, Australia}
\authorblockA{\authorrefmark{3}Australian Federal Police (AFP), Canberra, ACT 2600, Australia}
\authorblockA{\authorrefmark{4}Tata Consultancy Services (TCS) Ltd., Brisbane, QLD 4000, Australia}
}

\maketitle

\begin{abstract}
The Internet of Things (IoT) paradigm has displayed tremendous growth in recent years, resulting in innovations like Industry 4.0 and smart environments that provide improvements to efficiency, management of assets and facilitate intelligent decision making. However, these benefits are offset by considerable cybersecurity concerns that arise due to inherent vulnerabilities, which hinder IoT-based systems’ Confidentiality, Integrity, and Availability. Security vulnerabilities can be detected through the application of penetration testing, and specifically, a subset of the information-gathering stage, known as vulnerability identification. Yet, existing penetration testing solutions can not discover zero-day vulnerabilities from IoT environments, due to the diversity of generated data, hardware constraints, and environmental complexity.  Thus, it is imperative to develop effective penetration testing solutions for the detection of vulnerabilities in smart IoT environments. In this paper, we propose a deep learning-based penetration testing framework, namely  Long Short-Term Memory Recurrent Neural Network-Enabled Vulnerability Identification (LSTM-EVI). We utilize this framework through a novel cybersecurity-oriented testbed, which is a smart airport-based testbed comprised of both physical and virtual elements. The framework was evaluated using this testbed and on real-time data sources. Our results revealed that the proposed framework achieves about 99\% detection accuracy for scanning attacks, outperforming other four peer techniques.
\end{abstract}
\IEEEoverridecommandlockouts

%
\IEEEpeerreviewmaketitle

\section{Introduction}

The development of the Internet of Things (IoT) has resulted in a plethora of benefits, such as improving business processes, enhancing efficiency in resource management, and the introduction of the smart environment framework and Industry 4.0 \cite{Ruggeri2020,moustafa2021new}. By combining the functionality of individual IoT devices, new multi-level services can be defined and orchestrated, leading to innovative applications and intelligent decision-making systems \cite{Koroniotis2020}. However, the many advantages that the IoT offers are offset by significant cybersecurity concerns that arise from vulnerabilities inherently found in the sensors and actuators, impacting the confidentiality, integrity, and availability of the data and assets in the smart environments. With the number of deployed IoT devices rapidly rising, and zero-day vulnerabilities being increasingly exploited by attackers, it is crucial to address the challenge of securing the IoT \cite{9343133}.

Contemporary research and real-world attacks indicate that vulnerabilities render the IoT susceptible to a range of cyberattacks, due to firmware, hardware, protocol, authentication, or credential weaknesses \cite{Srivastava2020,AlHadhrami2021}. These vulnerabilities can be exploited by hackers for multiple purposes. Their end goals include modifying a device’s functionality by altering its internal state or the data it records, to steal information that is temporarily stored on an IoT device before it is forwarded to the gateway and the cloud backend, to cause malfunctions, and halt production lines, take control of the device and utilize it in large cyberattack campaigns, perform power depletion attacks by forcing a device to reply to repeated requests or utilize it as a springboard for lateral movement \cite{hussain2020machine}. Recent attacks, such as botnets and ransomware, have proven the destructive potential of IoT-powered cyberattacks, when routers, IP cameras, and other IoT devices were compromised, and orchestrated into launching immense volumetric DDoS attacks that were capable of disabling DNS servers and portions of the Internet \cite{Eustis2019}. It is, therefore, imperative to develop vulnerability detection methods that can be effectively applied to IoT infrastructure.

Penetration testing mechanisms can be generally classified into post-exploitation (passive), where the focus is on recovery and detection of ongoing attacks, and pre-exploitation (active), where tools take a proactive approach and periodically assess a network for vulnerabilities \cite{moustafa2018towards}. Penetration testing is an active method for assessing cybersecurity readiness and resilience when faced with a prepared attacker that persistently pursues the detection and exploitation of vulnerabilities. The process of penetration testing involves a team of experts that are actively attempting to breach the security of their target, using both software and hardware-specific tools for the detection and exploitation of weaknesses and vulnerabilities \cite{AlShebli2018}. The first crucial step during a penetration test or a cyberattack, and necessary action for vulnerability assessment, is the reconnaissance phase also known as information gathering. In the reconnaissance phase, scanning attacks are launched by specialized software that targets a remote host’s ports, seeking to identify the services that are associated with them and to detect any exploitable vulnerabilities \cite{Kumar2019}.

\textbf{Research Motivation}-- Contemporary security tools, for example, malware detection, intrusion detection and penetration testing, are designed to employ databases of pre-defined rules, signatures, and imposing policies for the detection of vulnerabilities. Although these tools are very accurate at detecting known vulnerabilities, they suffer from certain limitations \cite{Srivastava2020,moustaf2015creating}. Firstly, signature and rule-based detection tools can be circumvented by altering the behavior of an attack. Furthermore, they are incapable of detecting zero-day vulnerabilities \cite{moustafa2018towards,hussain2020machine}.  Consequently, there is a need to develop penetration testing tools and techniques for IoT settings, that can effectively detect vulnerabilities in live environments and overcome the limitations posed by conventional signature-based solutions \cite{Koroniotis2020,haider2020fgmc}. 

\textbf{Research Contribution}-- In this study, we present a novel deep learning-enabled penetration testing framework, named Long Short-Term Memory Recurrent Neural Network-Enabled Vulnerability Identification (LSTM-EVI). This framework focuses on the information gathering stage of penetration testing, and particularly on vulnerability detection. To develop and evaluate this framework, we use a novel cybersecurity-oriented testbed, which is a smart airport-based testbed that combines physical IoT devices and virtual elements. Through the smart airport testbed, both benign and malicious/scanning data were collected, transformed, and employed for the evaluation of the proposed framework. The main contributions of this paper are as follows:
\begin{itemize}
    \item We propose a new penetration testing framework that focuses on gathering information of vulnerabilities and learning scanning attacks.
    \item We designed and leveraged a smart airport-based testbed for the training and evaluation of the proposed framework.
\end{itemize}
The rest of this paper is organized as follows. Section II provides related work by discussing the current state of research on penetration testing in IoT settings. Section III focuses on the methodology for designing and constructing the proposed penetration testing framework. Section IV discussed the results obtained from experimenting with the proposed framework on the smart airport testbed that we developed. Finally, Section V provides the concluding remarks.

\section{Related Work}

Due to the growth of smart IoT environments, and the widespread adoption of these paradigms by organizations around the globe that incorporate them into their existing infrastructure, the security and integrity of these systems has become a popular topic of study \cite{Allakany2019,Abdalla2020,Yadav2020,Mikulskis2019}. In this section, we provide research related to the development and utilization of penetration testing techniques in IoT settings, focusing on vulnerability detection. For instance, Allakany \textit{et al.} \cite{Allakany2019} presented a plan for an end-to-end penetration testing framework for the IoT. The proposed framework was comprised of four modules, named planning, discovery, attack and reporting that were tasked with information gathering, vulnerability detection, vulnerability exploitation and reporting. The researchers provided an abstract overview of the proposed framework, and thus the vulnerability scanning methods were not discussed. Furthermore, the planning and discovery stages have overlapping functionality. 

Abdalla \textit{et al.} \cite{Abdalla2020} presented a penetration testing showcase for the IP camera “Onvif YY HD”. The proposed penetration testing process included 3 steps, starting with defining the area of investigation, implementation of process and outcome report presentation. The researchers employed several software tools found in Kali Linux machines, for network device discovery, network packet capture/analysis and cyber-attack implementation, that resulted in the identification of several vulnerabilities. Although this research delivered empirical data through penetration testing, they focused on a specific IP camera model, and thus generalisability is not ensured. Furthermore, the process was not automated and required the intervention of an investigator to be completed.

Yadev \textit{et al.} \cite{Yadav2020} introduced IoT-PEN, a client-server-based penetration testing framework for IoT environments. The researchers designed IoT-PEN to be automated and scalable, using target-graphs for defining potential attack plans and CVE IDs for accessing vulnerability information. The NVD database was utilized to detect vulnerabilities found in particular versions of software and devices. The state of IoT nodes is sent to the server in XML format, using the MQTT protocol, from where the NVD database was used to detect known vulnerabilities and provide recommendations. The provided work was evaluated on several network topologies, however, the researchers did not provide information about the IoT devices that were included in their testing environment. Furthermore, their implementation makes use of the NVD database for vulnerability detection, via utilizing the state of IoT nodes, thus missing zero-day vulnerabilities. Additionally, it is unclear if their proposed solution can be applied to IoT sensors/actuators in addition to gateways/aggregators, as the IoT-PEN requires to be installed and IoT sensors are constrained in resources, including their storage and processing capacity. 

Mikulskis \textit{et al.} \cite{Mikulskis2019} developed Snout, a software tool for identifying, scanning, attacking and assessing IoT devices using non-IP communication protocols. The proposed tool can perform penetration testing techniques and cyber attacks against IoT devices to assess their security. Snout’s penetration testing capabilities were evaluated using ZigBee and are capable of supporting multiple protocols that are commonly found in the IoT, such as BLE, Z-Wave, and LoRa. Its penetration testing capabilities include the evaluation of ZigBee-specific attacks and fuzzing. Although promising, this work does not specify what type of fuzzing the Snout tool utilizes. Furthermore, Snout’s capabilities were not showcased in a real or simulated scenario, thus its effectiveness in penetration testing and vulnerability detection has not been proven.

Although the presented studies \cite{Allakany2019,Abdalla2020,Yadav2020,Mikulskis2019} provided either a functional implementation or an extendable abstract design for a penetration testing tool tailor-made for IoT settings, the construction of a versatile and efficient deep learning-based penetration testing tool for smart airports that focuses on vulnerability detection, is still an unexplored topic. Additionally, some of the presented work either lacked any specific details of their testing environment or neglected to evaluate their solution through experimentation. Moreover, the penetration testing methods that were incorporated into the presented tools were, in some cases either not clarified, or reliant on existing tools that employ rule-based reasoning. This work seeks to address these shortcomings via developing a deep learning-based penetration testing framework, with its functionality focused primarily on vulnerability detection.

\section{ Mathematical Formulation of Penetration Testing}

To implement an automated deep learning-based penetration testing process, attacking and legitimate scenarios need to be executed. More importantly, for a system/network that is undergoing penetration testing, we define an initial condition ($I$), an attack activity set ($A$), the end state ($E$) and the attack probability set ($P$) \cite{moustafa2018towards}. The initial condition refers to the starting state of the system before the application of any attacks or penetration testing activities. The attack activity $(a \in A)$ corresponds to an action seeking to move from a previous state to a state closer to $E$, where $E$ denotes the exploitation of a vulnerability.

The attack probability $(p \in P)$ corresponds to a probability function that is given by the equation $f(p)=Prob(a/I,G)$, and gives the probability of an action leading from the initial condition to the end state. Essentially, automating the penetration testing process involves the identification of the states $S=[I,s_1,s_2,…,s_n,G]$, where the automated tool can transition using attack actions $A=[a_{I1},a_{I2},…,a{nn},a{nG}]$, where $n$ is the number of potential states that the system may assume. The automated penetration testing tool is tasked with identifying viable paths that will allow the transition from state $I$ to $G$, with the greatest probability of success $p$. If we define a transition function $F_{t}(InitialCondition,EndState,AttackAction)$, then the process can be viewed as: $F_{t_0}(I,G,0)=a_{I1},F_{t_1}(s_1,G,a_{I1})=a_{12},…,F_{t_n}(s_n,G,a_{(n-1)n}) =a_{nG}$

\section{ Proposed Deep learning-based Vulnerability Identification Framework}
In this section, we discuss the proposed deep learning-based penetration testing framework designed to launch reconnaissance-based scanning attacks and collect network traces. From the collected packet capture (pcap) files, network flows were extracted, pre-processed, enhanced through feature generation and labeled. The generated data was then employed for the training and testing of a deep learning model, that is designed to process network flows and detect vulnerabilities.

\begin{figure*}[ht]
\centering
    \includegraphics[width=0.85\textwidth]{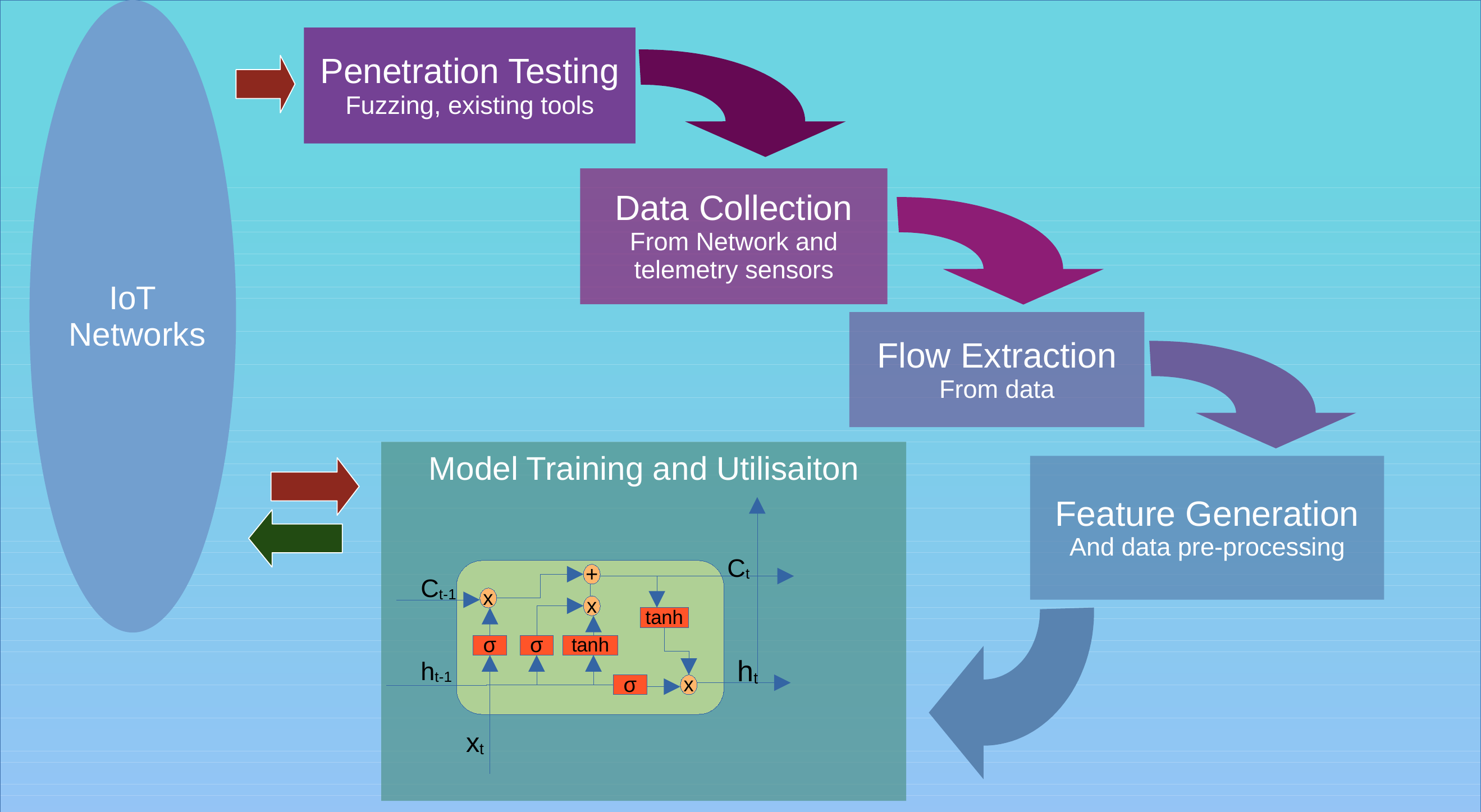}
    \caption{\label{fig:Detection_Framework} Proposed framework of vulnerability identification, which is a case study of implementation in smart environments, such as smart airports}
\end{figure*}

In this work, a vulnerability identification framework is presented to incorporate the data collection, processing and application for the detection of Reconnaissance scanning attacks. As shown in Figure \ref{fig:Detection_Framework}, there are five main components of this framework. Each of which is distinct in its functionality and provides services to the next components, as explained in the following items.
\begin{itemize}
    \item \textbf{Penetration Testing} - In this component, the penetration testing process is initiated through the use of tools, such as Nessus, Scapy and Zeek. They employ fuzzing and rule-based techniques that scan the network for vulnerabilities. 
    \item \textbf{Data Collection} - In this component, network traffic and telemetry data obtained from the testbed's network and sensors respectively, are collected.
    \item \textbf{Flow Extraction} - In this component, the collected data is processed and network flows are extracted to form a prime dataset that will be used for training and evaluation of a DL model.
    \item \textbf{Feature Generation} - In this component, the prime dataset is further analyzed and enhanced, by generating new descriptive features that will enhance the performance of the DL model. The generated dataset is designed.
    \item \textbf{Model Training and Utilisation} - In this component, the generated dataset is utilized for the training and validation of a DL LSTM-based model that is used for vulnerability identification.
\end{itemize}

\subsection{Testbed configuration for penetration testing purposes}

A testbed was designed to represent a realistic smart airport environment and evaluate the fidelity of deep learning-based penetration testing systems.  The testbed incorporates physical IoT devices, gateways, switches, computers, and virtual machines configured to interact in several attacking and legitimate scenarios. The design of such a testbed, which includes both physical IoT devices as well as virtual machines has certain advantages. To begin with, the physical off-the-shelf IoT devices can be found in real-world smart environments and thus ensure correct representation. Furthermore, unlike virtualized IoT devices, their physical counterparts allow the exploitation of firmware and hardware vulnerabilities, which need to be represented in the collected data, as it is crucial for the development of deep learning-based vulnerability analysis tools.

The inclusion of VMs renders our testbed versatile, portable, and configurable, as VMs can be easily replaced, moved between computers and their resources adjusted according to circumstances. Additionally, a data management system was developed along with the testbed, consisting of a telemetry-based module that is tasked with the collection of telemetry data that the sensors produce, and a network traffic analysis module that handles the collection of packets, the extraction of network flows, the generation of new features and correct labelling of both network and telemetry data.

The testbed consists of three zones; each of which represents a unique space in a smart airport is assigned a single gateway with USB NICs for the four wireless communication protocols, Bluetooth, ZigBee, Z-Wave and Lora of the included sensors, and several sensors that are indicative of the location represented by the zone. The gateways are connected to an access point that facilitates local communications, through WiFi, between the VMs and the rest of the testbed. The telemetry data that the sensors produce is published to an MQTT broker, through open-source software, from which it can be retrieved and analyzed. The testbed includes 20 VMs 10 of which (benign VMs) interact with each other and the zones and perform data collection, processing, management and visualization. The rest are tasked with targeting the sensors, gateways and the benign VMs with cyber-attacks (attacking VMs), to represent the attacking scenarios when the testbed is employed for the construction of datasets.

\begin{figure*}
\centering
    \includegraphics[width=0.85\textwidth]{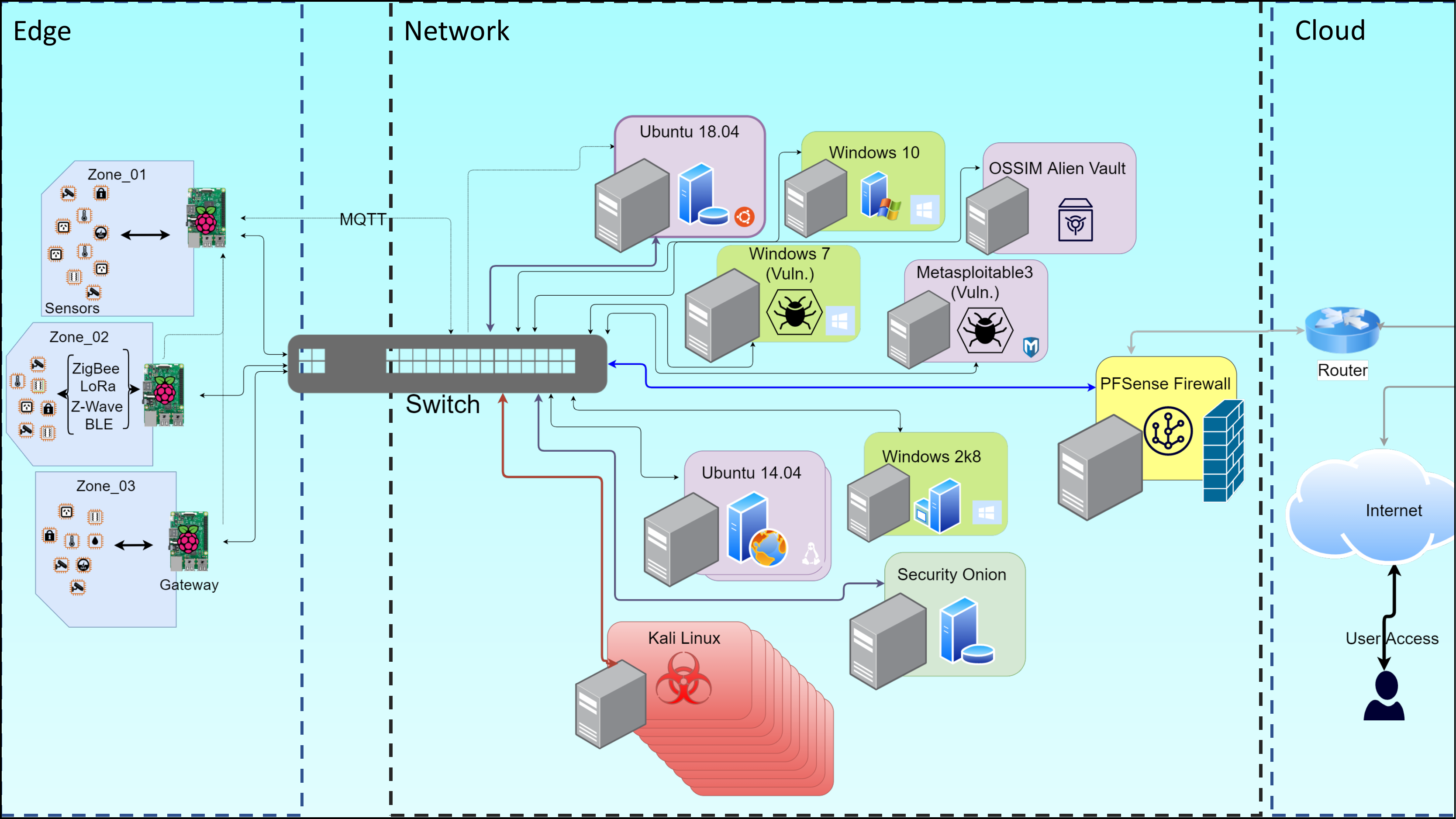}
    \caption{\label{fig:Smart_Airport_Testbed_Architecture} Proposed Smart Airport Testbed Architecture}
\end{figure*}

Various physical IoT devices were attached to the testbed, and they included environmental sensors (humidity, temperature, air pressure), contact sensors, motion sensors, control devices such as switches, condensation sensors and IP cameras. This testbed was designed to collect and refine data, produce representative cybersecurity-based IoT datasets, and develop/validate efficient cyber-defence IoT-compatible solutions, such as vulnerability analysis tools.  The proposed smart airport testbed's architecture is presented in Figure \ref{fig:Smart_Airport_Testbed_Architecture}.

To maximize the probability of success in their activities, cyber attackers and penetration testers alike often choose to employ established methods and strategies, when planning their attacks. The cyber kill chain model, developed by Lockheed Martin \cite{Dargahi2019}, analyses the necessary steps that Advanced Persistent Threats (APTs) which can be an attacker or a penetration tester, need to consider when attempting to attack and compromise a target in a network. The framework includes seven stages; each of which provides some information to the next, facilitating the successful infiltration of a computer and its resources. The stages of  cyber kill chain consist of Reconnaissance, Weaponisation, Delivery, Exploitation, Installation, Command and Control, Actions on Objectives. 

In the Reconnaissance stage, the attackers gather as much information as possible about their target, seeking to identify a weakness they can use to their advantage. For example, a penetration tester, may launch scanning and probing attacks during the reconnaissance stage, to identify Internet-facing machines, their open ports and the services they provide. By doing so, they can potentially detect vulnerable services, with known exploitation. In the Weaponisation stage, the attackers use the vulnerabilities detected in the Reconnaissance stage to design their attack exploit, by combining the exploit code with a backdoor in a payload.

In the Delivery and Exploitation stages, the crafted payload is sent to the target in the selected form (e-mail, USB stick, packet sequence, etc.) and the exploit is executed. In the Installation and Command and Control (C\&C) stages, the exploit that was executed previously facilitates the installation of malware that, once correctly installed, causing the backdoor code to run and establish a reverse channel with the attacker, initiating a C\&C channel. In the final stage, after the attackers have established a secure and reliable channel of communication with the compromised machine, they can proceed with their goal, for instance, to exfiltrate data, perform lateral movement and more. In this paper, we focus primarily on the Reconnaissance stage, as it is during that stage that vulnerabilities in IoT devices and other networked entities can be detected.

Before any vulnerabilities can be detected, the first action that needs to be performed in the Reconnaissance stage is the detection of live hosts in the targeted network. For that purpose, we utilised the Nmap tool to perform a ping-based scan and an initial service detection, which resulted in a list of available hosts that would be later targeted with more sophisticated scanning-based attacks. The command that we employed for host detection was:
\[nmap -sV --script=banner\  Target\_IP\_Address\]
with the $-sV$ specifying that nmap should attempt to identify the version of services that it detects, and the $--script=banner$ enables the banner grabbing functionality, that captures the banners that services advertise, with banners being the initial text that the services generate upon receiving a request through their respective ports. 

Regarding the underlying mechanics that are employed for the detection of vulnerabilities in the networks and their connected devices, two primary categories exist, rules/signature-based and machine or deep learning-based \cite{Li2019, Sarker2020}. Many commercial and open-source tools such as Nessus, Zeek and Scapy, rely on rules and signatures crafted through expert knowledge that results in the creation of static profiles/signatures of vulnerabilities, which are detected if the response of carefully crafted network probe matches the signature or the rule.

\subsection{ Proposed Long  Short-Term  Memory  Recurrent  Neural  Network-Enabled Vulnerability Identification Model}

Deep Learning (DL)-based penetration testing tools would scan the network and process collected data, to detect patterns that indicate the existence of a vulnerability, based on the output of the DL model that has been trained on curated and representative data, allowing them to generalize well and be more versatile. To test for vulnerabilities, one technique that has been employed in the past in multiple scenarios is fuzzing \cite{ Wang2020}. Fuzzing vulnerability testing relies on the generation of pseudo-random data that is forwarded to a target machine, and if that elicits a legitimate response from the system, then a vulnerability has been detected. 

We employed Nessus, Zeek and Scapy to perform fuzzing scanning attacks against the network-enabled components of the testbed, which resulted in the generation of network traffic that was collected, processed, and labelled. The generated data was utilized to train a deep learning model, namely a Long Short-Term Memory Recurrent Neural Network (LSTM-RNN), on 1-timestep intervals, that would output a class feature as either being “0” indicating normal traffic or “1” indicating a scanning attack.

Deep Learning is a subclass of neural networks, designed to have multiple layers and hundreds of units \cite{Zaccone2017}, resulting in models that perform well when processing large volumes of data. The LSTM that is employed in the LSTM-EVI framework, is a version of the RNN that overcomes the latter’s limitations when tasked with learning long sequences of data, by employing certain mathematical constructs, known as gates, that control the model’s memory.  

An LSTM model that processes data with multiple timesteps, receives the first input data, processes it through the LSTM cell by applying weights and activation gates such as sigmoid and tanh, and forwards the hidden state to higher layers of the network, passing the computed cell and hidden states to the next timestep. There are 3 gates in an LSTM that manage the cell state and, in turn, affect the hidden state.

The forget gate, given in Equation \ref{eq:forget_gate}, determines what portion of the previous cell state is going to be forgotten, by diminishing those values to $0$.  The input gate, given in Equation \ref{eq:input_gate}, determines what parts of the cell state will be modified, and specifically what information will be added. Equation \ref{eq:new_cell_state} provides the new cell state that is calculated by combining the previous cell state, the information gate (Equation \ref{eq:input_gate}) and the new information ( Equation \ref{eq:new_information}). 

The new hidden state, which is the output of an LSTM cell, is forwarded to the next cell when working with multiple timesteps, is determined by the output gate, as formulated in Equation \ref{eq:output_gate}. During training, we selected binary cross-entropy loss, as it performs well when tasked with training binary classifiers. The cost in a mini-batch during training is given in Equation \ref{eq:cross_entropy}.

\begin{equation}\label{eq:forget_gate}
    f_t=\sigma(w_f*[h_{t-1},x_t]+b_f, f_t \in [0,1]
\end{equation}
\begin{equation}\label{eq:input_gate}
    i_t=\sigma(w_i*[h_{t-1},x_t]+b_i, i_t \in [0,1]
\end{equation}
\begin{equation}\label{eq:new_information}
    C’_t=\tanh(w_c*[h_{t-1},x_t]+b_c,  C’_t \in [-1,1]
\end{equation}
\begin{equation}\label{eq:new_cell_state}
    C_t=f_t*C_{t-1}+i_t*C’_t
\end{equation}
\begin{equation}\label{eq:output_gate}
    h_t=\sigma(w_o*[h_{t-1},x_t]+b_o)*\tanh(C_t)
\end{equation}

\begin{equation}\label{eq:cross_entropy}
    C=-\frac{1}{n}\sum^n_1[y_ilog(y'_i)+(1-y_i)log(1-y'_i)]
\end{equation}

 \[\forall y,y' \in [0,1], i \in [1,n] \]

\section{Experimental Results}
 In this section, the proposed penetration testing framework is evaluated and compared with five machine learning models on data acquired from the smart airport testbed for detecting scanning attacks. The models that were evaluated are LSTM, Multi-layered Perceptron (MLP), Support Vector Machine (SVM), Naive Bayes (NB), and K-Nearest Neighbour (KNN).
 
\subsection{Environment Setup}

The process of setting up the smart airport testbed involved connecting IoT sensors that utilise different wireless communication protocols (e.g., BLE, Z-Wave, ZigBee, Lora, and WiFI), with coordinator devices (Raspberry Pi) and the rest of the testbed  (VMs, Switches, Routers, Wireless Access Points). Data was collected by the Tap machine (Ubuntu 18.04) and the scanning attacks were launched by the Kali VMs that can be seen in Figure \ref{fig:Smart_Airport_Testbed_Architecture}. After the dataset was finalised, two DL models were trained in an Ubuntu VM with Intel Core i7-10510U @ 1.8 GHz processor and 3 GB RAM. The programming environment that was used was Python 3.8, with the Keras and Tensorflow packages selected for defining, training and evaluating the DL models.  

\subsection{Dataset collected from the testbed}

From the initial pcap files that were collected from the testbed, data features were extracted to train and test machine learning for classifying scanning attacks.  Network flow features were then converted to a numerical form, using label encoding, with the process displayed in Algorithm \ref{algo:convert_numerical_categorical}. Next, to improve the performance of the DL models, a min-max normalisation function was applied to all features (excluding the class feature), transforming their values to be within the range $[0,1]$. This range would allow deep learning models to effectively learn legitimate and scanning observations without bias towards a particular class. 

The generated dataset was split into training (70\%) and testing (30\%) subsets, a common training/texting split that has been shown to improve results, and for the LSTM model, the subsets were modified to include a single timestep. The structure and hyperparameters of the two DL models can be seen in Table \ref{tab:DL_MDS_Hyper}. The structure of each model includes the number of features as the first layer(16), and for the LSTM-RNN, we defined the first two hidden layers (64 and 128 units) as LSTM, stacking five dense layers to perform classification. To enable reproducibility, we explicitly set the random seeds for TensorFlow and Numpy to be $22$.

\begin{algorithm}[h]
\SetAlgoLined
 dataset=Load\_Dataset()\;
 \For{column in dataset.columns()}{
    \uIf{not column.is\_numeraical()}{
        i=0\;
        \For{value in column.unique\_values()}{
            column.exchange(value,i)\;
            i+=1\;
        }
        dataset.update(column)\;
    }
    
 }
 return dataset\;
 
 \caption{\label{algo:convert_numerical_categorical}Encoding of categorical values to numerical}
\end{algorithm}

\begin{table*}[tp]
\caption{Deep Learning Models' Hyperparameters}
    \label{tab:DL_MDS_Hyper}
    \centering
    \begin{tabular}{|p{0.16\linewidth}|p{0.22\linewidth}|p{0.06\linewidth}|p{0.1\linewidth}|p{0.1\linewidth}|p{0.08\linewidth}|}
    \hline
         Model&Structure&Optimiser&Learning Rate&Batch Size&Epochs  \\\hline
         MLP&[16,32,128,512,32,1]&Adam&0.001&560&10\\\hline
         LSTM-RNN&[16,64,128,256,512,64,64,32,1]&Adam&0.001&560&10\\\hline
    \end{tabular}
\end{table*}

\subsection{Results and Discussions}

\begin{figure}
    \centering
    \includegraphics[width=0.48\textwidth]{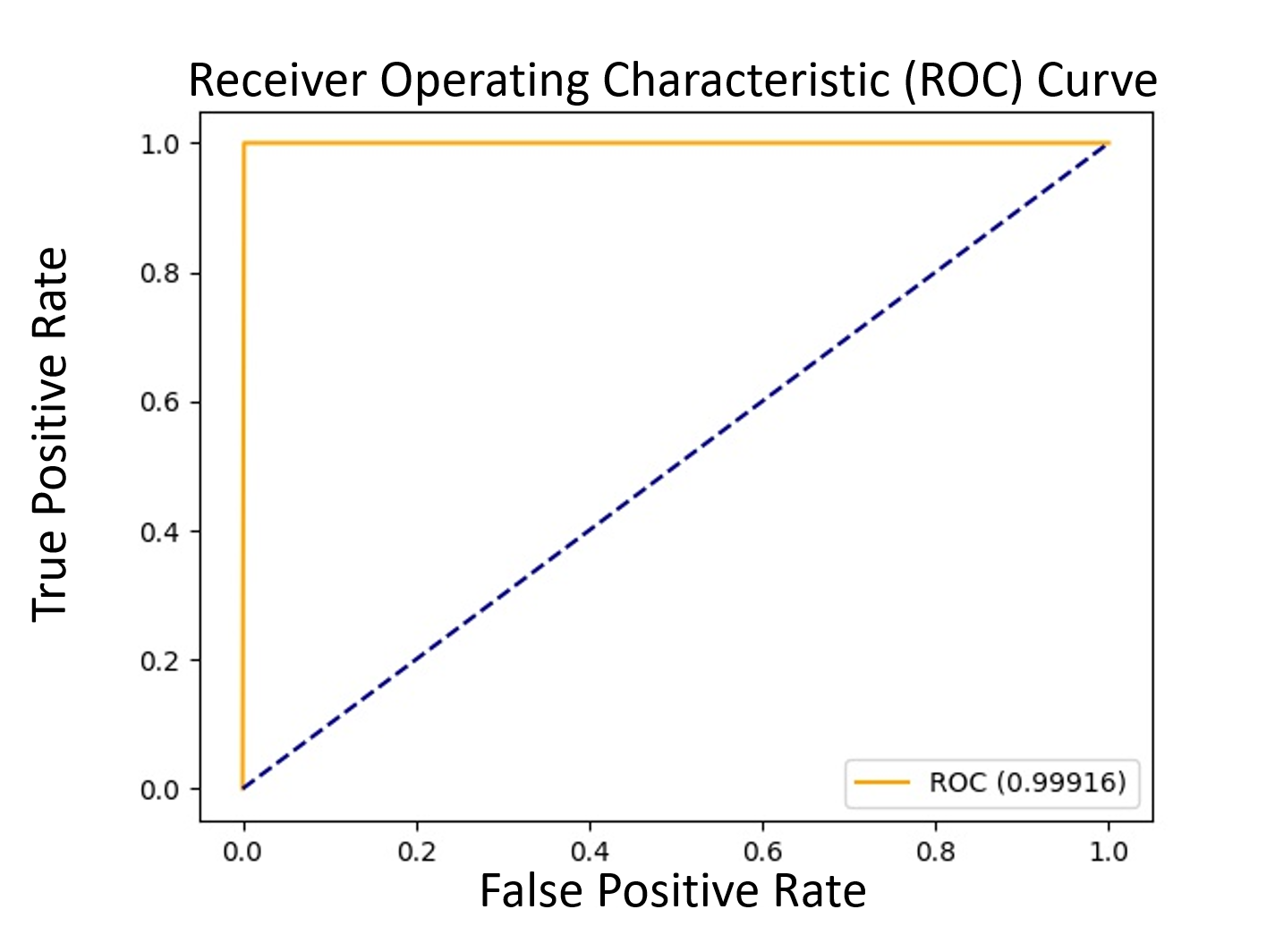}
    \caption{\label{fig:LSTM_ROC}ROC Curve for LSTM model.}
\end{figure}

Each deep learning model was trained for 10 epochs, using the Adam optimiser on the same batch size. The ML/DL models were trained to perform binary classification, returning either $0$ for normal flows or $1$ for attack instances. The decision threshold, which determines if the predicted class feature is either $0$ or $1$ for both models was set at $0.5$. The confusion matrix of the LSTM model can be seen in Table \ref{tab:LSTM_Confusion} and its Receiver Operating Characteristics (ROC) curves in Figures \ref{fig:LSTM_ROC}. Additional metrics that were calculated to compare the performance of the aforementioned models, including Accuracy, Precision, Recall, Specificity and F-score are provided in Table \ref{tab:Models_Metrics}.

Of all the tested models, the LSTM outperformed the rest, achieving  $0.9991$ classification accuracy. MLP, SVM and KNN displayed similar performance, with the Naive Bayes model having the worst classification accuracy at just over $0.5$. As can be seen, by the calculated metrics, both models achieved high accuracy and low error rates. The reason for the LSTM’s performance is slightly better than is that LSTM is training more weights due to the added forget and state gates.  In our work, we transformed our data to have only 1 timestep, thus the hidden states and cell states that help the LSTM to maintain the memory of long sequences of data was not used effectively. Nevertheless, these results indicate that the LSTM can be used effectively to detect scanning attacks, the first step in the cyber kill chain, and an indication that an APT has started to target a system.

\begin{table}[h]
\centering
\caption{Confusion Matrix for LSTM-RNN}
    \label{tab:LSTM_Confusion}
    \begin{tabular}{|c|c|c|}
    \hline
         Prediction/Actual&Positive&Negative  \\\hline
         Positive&605&1\\\hline
         Negative&0&594\\\hline
    \end{tabular}
\end{table}

\begin{table}[h]
\centering
\caption{Performance Metrics for the two Deep Learning Models.}
    \label{tab:Models_Metrics}
    \begin{tabular}{|c|c|c|c|c|c|}
    \hline
         Metric&Naive Bayes&KNN&SVM&MLP&\textbf{LSTM}\\\hline
         Accuracy&0.5241&0.9983&0.9983& 0.9983& 0.9991 \\\hline
         Precision&0.5144&0.9967&0.9967& 0.9967& 0.9983  \\\hline
         Recall&0.9981 &1.0&1.0& 1.0& 1.0 \\\hline
         Specificity&0.042&0.9966&0.9966& 0.9966& 0.99831 \\\hline
         F-score&0.679&0.9983&0.9983&0.9983 & 0.9991 \\\hline
         AUC&0.5201&0.9983&0.9983& 0.9983& 0.9991 \\\hline

    \end{tabular}
\end{table}

\section{Conclusion}
This study has presented a deep learning-enabled penetration testing framework for identifying vulnerabilities in IoT networks, such as smart airports. Specifically, an LSTM-RNN model was developed for detecting scanning attacks in an IoT network-based smart airport testbed. Scanning attacks are part of Reconnaissance, the first stage of the cyber kill chain model. To collect data and construct a dataset for this work, we utilised developed a smart airport hybrid testbed that incorporated both physical IoT devices and virtual machines. For the scanning attacks, we focused on vulnerability detection through fuzzing and penetration testing tools, such as Nmap, Zeek, Nessus and Scapy. A network dataset was extracted from the testbed for evaluating the performances of DL models. Our experiments indicated that the proposed framework can achieve high performance in detecting scanning attacks compared with other compelling techniques. In the future, we intend to generate a large-scale dataset that will incorporate a greater range of scanning attacks. Furthermore, we intend to investigate the applicability of reinforcement learning in detecting vulnerabilities in IoT settings.

\section*{Acknowledgment}
The work was supported in part by the Cyber Security Research Centre Ltd., funded by the Australian Government’s Cooperative Research Centres Programme under Grant RG201120.
\bibliography{vulnan}

\begin{thebibliography}{10}

\bibitem{Ruggeri2020}
G.~Ruggeri, V.~Loscrí, M.~Amadeo, and C.~T. Calafate, ``The internet of things
  for smart environments,'' {\em Future Internet}, vol.~12, no.~3, 2020.

\bibitem{moustafa2021new}
N.~Moustafa, ``A new distributed architecture for evaluating ai-based security
  systems at the edge: Network {TON\_IoT} datasets,'' {\em Sustainable Cities
  and Society}, p.~102994, 2021.

\bibitem{Koroniotis2020}
N.~{Koroniotis}, N.~{Moustafa}, F.~{Schiliro}, P.~{Gauravaram}, and
  H.~{Janicke}, ``A holistic review of cybersecurity and reliability
  perspectives in smart airports,'' {\em IEEE Access}, vol.~8,
  pp.~209802--209834, 2020.

\bibitem{9343133}
N.~Moustafa, M.~Keshk, E.~Debie, and H.~Janicke, ``Federated {TON\_IoT} windows
  datasets for evaluating ai-based security applications,'' in {\em 2020 IEEE
  19th International Conference on Trust, Security and Privacy in Computing and
  Communications (TrustCom)}, pp.~848--855, 2020.

\bibitem{Srivastava2020}
A.~Srivastava, S.~Gupta, M.~Quamara, P.~Chaudhary, and V.~J. Aski, ``Future
  {IoT}-enabled threats and vulnerabilities: State of the art, challenges, and
  future prospects,'' {\em International Journal of Communication Systems},
  vol.~33, no.~12, p.~e4443, 2020.
\newblock e4443 IJCS-19-0930.R3.

\bibitem{AlHadhrami2021}
Y.~Al-Hadhrami and F.~K. Hussain, ``{DDoS} attacks in {IoT} networks: a
  comprehensive systematic literature review,'' {\em World Wide Web},
  pp.~1--31, 2021.

\bibitem{hussain2020machine}
F.~Hussain, R.~Hussain, S.~A. Hassan, and E.~Hossain, ``Machine learning in
  {IoT} security: Current solutions and future challenges,'' {\em IEEE
  Communications Surveys \& Tutorials}, vol.~22, no.~3, pp.~1686--1721, 2020.

\bibitem{Eustis2019}
A.~G. Eustis, ``The {Mirai} {Botnet} and the importance of {IoT} device
  security,'' in {\em 16th International Conference on Information
  Technology-New Generations (ITNG 2019)} (S.~Latifi, ed.), (Cham), pp.~85--89,
  Springer International Publishing, 2019.

\bibitem{moustafa2018towards}
N.~Moustafa, B.~Turnbull, and K.-K.~R. Choo, ``Towards automation of
  vulnerability and exploitation identification in {IIoT} networks,'' in {\em
  2018 IEEE International Conference on Industrial Internet (ICII)},
  pp.~139--145, IEEE, 2018.

\bibitem{AlShebli2018}
H.~M.~Z. Al~Shebli and B.~D. Beheshti, ``A study on penetration testing process
  and tools,'' in {\em 2018 IEEE Long Island Systems, Applications and
  Technology Conference (LISAT)}, pp.~1--7, IEEE, 2018.

\bibitem{Kumar2019}
R.~Kumar and K.~Tlhagadikgora, ``Internal network penetration testing using
  free/open source tools: Network and system administration approach,'' in {\em
  Advanced Informatics for Computing Research} (A.~K. Luhach, D.~Singh, P.-A.
  Hsiung, K.~B.~G. Hawari, P.~Lingras, and P.~K. Singh, eds.), (Singapore),
  pp.~257--269, Springer Singapore, 2019.

\bibitem{moustaf2015creating}
N.~Moustaf and J.~Slay, ``Creating novel features to anomaly network detection
  using darpa-2009 data set,'' in {\em Proceedings of the 14th European
  Conference on Cyber Warfare and Security. Academic Conferences Limited},
  pp.~204--212, 2015.

\bibitem{haider2020fgmc}
W.~Haider, N.~Moustafa, M.~Keshk, A.~Fernandez, K.-K.~R. Choo, and A.~Wahab,
  ``Fgmc-hads: Fuzzy gaussian mixture-based correntropy models for detecting
  zero-day attacks from linux systems,'' {\em Computers \& Security}, vol.~96,
  p.~101906, 2020.

\bibitem{Allakany2019}
A.~Allakany, G.~Yadav, V.~Kumar, K.~Paul, K.~Okamura, and C.~Center, ``An
  automated end-to-end penetration testing for {IoT},'' {\em The National
  Conference of the Information Processing Society}, vol.~5, p.~05, 2019.

\bibitem{Abdalla2020}
P.~A. Abdalla and C.~Varol, ``Testing {IoT} security: The case study of an ip
  camera,'' in {\em 2020 8th International Symposium on Digital Forensics and
  Security (ISDFS)}, pp.~1--5, IEEE, 2020.

\bibitem{Yadav2020}
G.~Yadav, K.~Paul, A.~Allakany, and K.~Okamura, ``{{IoT-PEN}: An E2E
  Penetration Testing Framework for {IoT}},'' {\em Journal of Information
  Processing}, vol.~28, pp.~633--642, 2020.

\bibitem{Mikulskis2019}
J.~Mikulskis, J.~K. Becker, S.~Gvozdenovic, and D.~Starobinski, ``{Snout: An
  Extensible IoT Pen-Testing Tool},'' in {\em Proceedings of the 2019 ACM
  SIGSAC Conference on Computer and Communications Security}, pp.~2529--2531,
  2019.

\bibitem{Dargahi2019}
T.~Dargahi, A.~Dehghantanha, P.~N. Bahrami, M.~Conti, G.~Bianchi, and
  L.~Benedetto, ``A cyber-kill-chain based taxonomy of crypto-ransomware
  features,'' {\em Journal of Computer Virology and Hacking Techniques},
  vol.~15, no.~4, pp.~277--305, 2019.

\bibitem{Li2019}
Z.~{Li}, D.~{Zou}, J.~{Tang}, Z.~{Zhang}, M.~{Sun}, and H.~{Jin}, ``A
  comparative study of deep learning-based vulnerability detection system,''
  {\em IEEE Access}, vol.~7, pp.~103184--103197, 2019.

\bibitem{Sarker2020}
I.~H. Sarker, Y.~B. Abushark, F.~Alsolami, and A.~I. Khan, ``Intrudtree: a
  machine learning based cyber security intrusion detection model,'' {\em
  Symmetry}, vol.~12, no.~5, p.~754, 2020.

\bibitem{Wang2020}
Y.~Wang, P.~Jia, L.~Liu, C.~Huang, and Z.~Liu, ``A systematic review of fuzzing
  based on machine learning techniques,'' {\em PloS one}, vol.~15, no.~8,
  p.~e0237749, 2020.

\bibitem{Zaccone2017}
G.~Zaccone, M.~Karim, and A.~Menshawy, {\em Deep Learning with TensorFlow}.
\newblock Packt Publishing, 2017.

\end{thebibliography}
\bibliographystyle{ieeetr}

\end{document}